\documentclass{article}
\usepackage{spconf,amsmath,graphicx,hyperref}
\usepackage{multirow}
\usepackage{booktabs}
\usepackage{graphicx}
\usepackage{amssymb} 
\usepackage{bbding,pifont,tikz}
\usetikzlibrary{arrows,positioning}
\usepackage{colortbl}
\usepackage[table,svgnames]{xcolor}

\title{Rethinking the joint estimation of magnitude and phase for time-frequency domain neural vocoders}
\name{Lingling Dai$^{\star\ast}$, Andong Li$^{\star \ast}$, Tong Lei$^{\dagger}$, Meng Yu$^{\ddagger}$, Xiaodong Li$^{\star\ast}$, Chengshi Zheng$^{\star\ast}$}
\address{
	${\star}$ Institute of Acoustics, Chinese Academy of Sciences, Beijing, China\\
	$\dagger$Tencent AI Lab, Shen Zhen, China\\
	$\ddagger$Tencent AI Lab, Bellevue, WA, USA\\
	$^{\ast}$University of Chinese Academy of Sciences, Beijing, China}
\begin{document}
	\ninept
	\maketitle
	\begin{abstract}
		Time-frequency (T-F) domain-based neural vocoders have shown promising results in synthesizing high-fidelity audio. Nevertheless, it remains unclear on the mechanism of effectively predicting magnitude and phase targets jointly. In this paper, we start from two representative T-F domain  vocoders, namely Vocos and APNet2, which belong to the single-stream and dual-stream modes for magnitude and phase estimation, respectively. When evaluating their performance on a large-scale dataset, we accidentally observe severe performance collapse of APNet2. To stabilize its performance, in this paper, we introduce three simple yet effective strategies, each targeting the topological space, the source space, and the output space, respectively. Specifically, we modify the architectural topology for better information exchange in the topological space, introduce prior knowledge to facilitate the generation process in the source space, and optimize the backpropagation process for parameter updates with an improved output format in the output space. Experimental results demonstrate that our proposed method effectively facilitates the joint estimation of magnitude and phase in APNet2, thus bridging the performance disparities between the single-stream and dual-stream vocoders. 
		
	\end{abstract}
	\section{Introduction}
	\label{sec:intro}
	\vspace{-0.2cm}
	Neural vocoders aim to reconstruct audible waveforms from intermediate acoustic features or hidden representations using deep neural networks (DNNs)~{\cite{vocos}}. Driven by the escalating demands of audio-related applications, they have garnered substantial attention and achieved remarkable progress in recent years~\cite{seedtts,liu2024audiosr,liu2021voicefixer,hifi++,li2025neural,bigvgan}. Existing methods can be roughly categorized into time-domain and time-frequency (T-F) domain approaches, where the latter have gained increasing prominence due to the appealing advantages in perceptual quality and inference efficiency~\cite{istftnet,lightvoc,freev}.
	
	Unlike consecutive upsampling operations, T-F domain-based neural vocoders typically estimate the real and imaginary (RI) parts or magnitude-phase (MP) pairs, followed by the inverse short-time Fourier transform (iSTFT) for target waveform generation. Therefore, a core challenge lies in \textit{how to effectively predict magnitude and phase targets jointly?} When inspecting related audio processing tasks \emph{e.g.}, speech enhancement (SE) and bandwidth extension (BWE), we notice that two typical architectural topologies are usually adopted: \textit{single-stream} and \textit{dual-stream}. For the former, magnitude and phase share most deep modeling units, with separate output heads for dual-target prediction~{\cite{PHASEN,rndvoc,baenet,gcrn,ICRN,wang2023tf,QUICKVC}}. For the latter, by contrast, dual-stream architectures employ independent modeling streams with minimal or no inter-stream interactions for MP prediction~\cite{apnet2,AP-BWE,ai2023apnet}. Nonetheless, it still remains an open question to determine an optimal architecture modality.
	
\begin{figure}[t]
		\centering
		\includegraphics[width=0.46\textwidth]{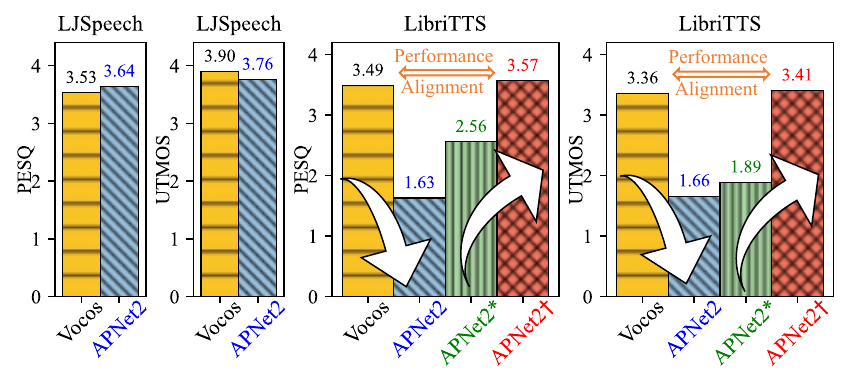}
		\vspace{-18pt}
		\caption{The performance illustration of Vocos and APNet2 on the LJSpeech dataset and LibriTTS dataset, where APNet2* is aligned with Vocos in implementation details and APNet2$\dag$ denotes our improved version.}
		\label{fig:pesq_comparison}
		\vspace{-0.5cm}
	\end{figure}

	When switching back to the vocoder task, we compare two representative models: Vocos~\cite{vocos} and APNet2~{\cite{apnet2}}. The reasons are two-fold. First, they belong to the single-stream and dual-stream, respectively. Besides, both of them utilize the ConvNext block~{\cite{liu2022convnet}} and its variant, \emph{i.e.}, ConvNext v2~{\cite{woo2023convnext}}, as the basic modeling unit. We evaluate their performance on two benchmarks: LJSpeech~{\cite{ljspeech}} and LibriTTS~{\cite{libritts}}, and PESQ and UTMOS scores are shown in Fig.~{\ref{fig:pesq_comparison}}. Surprisingly, while both models exhibit comparable reconstruction quality on the single-speaker LJSpeech dataset, APNet2 suffers from significant performance collapse on LibriTTS, a larger benchmark with diverse acoustic recordings. To investigate the root, we first adjust APNet2 to align with Vocos in terms of basic modeling unit and training configurations, yielding APNet2$^{*}$, however, a large performance gap still exists{\footnote{Similar performance gap is observed in other large-scale datasets, \emph{e.g.}, Libriheavy dataset~{\cite{kang2024libriheavy}}.}}. Therefore, we conjecture the ``dual-stream design'' to be the primary reason to cause the performance collapse in the vocoder task. Further, for more diverse generation scenarios, due to the inherent wrapping property of phase, in the existing dual-stream architecture modality, the phase branch may lack \textbf{adequate} feature guidance from the magnitude branch{\footnote{Although some feature interactions strategies can be introduced for guidance, we observe marginal improvements, as presented in Table~{\ref{tab:modality_results}}.}}, which seems especially significant for waveform reconstruction in the T-F domain.

	\begin{figure*}[t]
		\centering
		\includegraphics[width=0.75\textwidth]{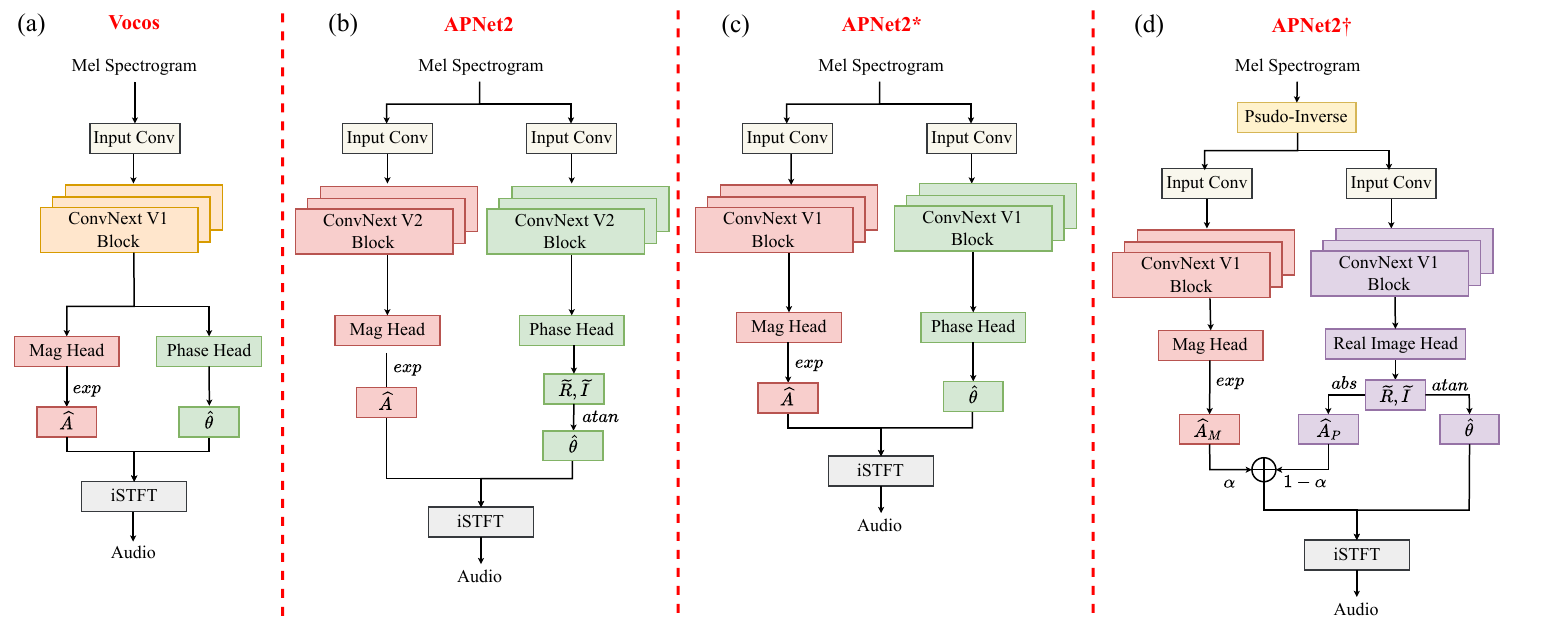}
		\vspace{-10pt}
		\caption{The model structure illustrations of Vocos, APNet2, APNet2* and APNet2$\dag$.}
		\label{fig:framework}
		\vspace{-13pt}
	\end{figure*}
	To this end, we propose to resolve the performance collapse observed in existing ``dual-stream'' based neural vocoders with three simple yet effective strategies, each targeting the \textbf{topological space}, \textbf{source space}, and \textbf{output space}. Specifically, for the topological space, we propose a straightforward and intuitive strategy by modifying the architectural topology with more information interaction and weight sharing between the magnitude and phase streams. For the source space, when the compressed Mel spectrum is considered as the input, it may be unsuitable to serve as the information source, as it is misaligned with the magnitude spectrum at the T-F bin level. Motivated by~{\cite{freev,rndvoc}}, we project the Mel spectrum into the range space of the target spectrum by applying the pseudo-inverse Mel filter, providing a more aligned and informative input for joint MP prediction. For the output space, instead of direct phase prediction or its trigonometric representations, we enforce the phase branch to generate partial magnitude components, enabling cross-stream   
	magnitude optimization from the magnitude branch. This mechanism can establish implicit feature dependency between the two streams. Comprehensive results demonstrate that our proposed strategies effectively mitigate the performance collapse in APNet2. Our contributions are summarized as follows:\\
	\noindent \ding{113}~(1) We conduct quantitative experiments to investigate the performance disparities across different architectural topologies, taking Vocos and APNet2 as a typical example.\\
	\noindent \ding{113}~(2) We reformulate the generation and optimization process in magnitude and phase modeling from three distinct spaces.\\
	\noindent \ding{113}~(3) We conduct extensive experiments to validate the effectiveness of the proposed strategies, which effectively improve the model performance on joint magnitude and phase estimation.
	\vspace{-8pt}
	
	\begin{figure}[h]
		\centering
		\includegraphics[width=0.48\textwidth]{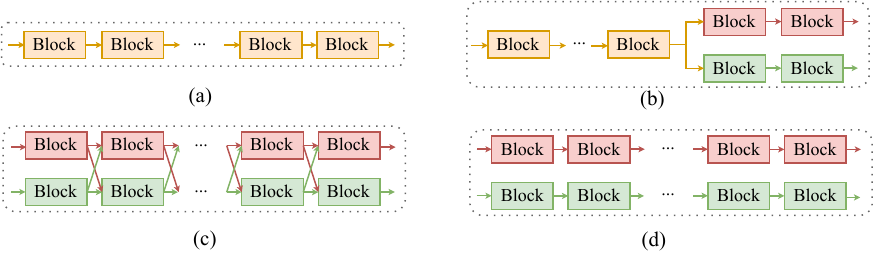}
		\vspace{-12pt}
		\caption{Overall diagram of different structural designs. (a) Shared structure between magnitude and phase estimations. (b) Partially shared structure between magnitude and phase estimations. (c) Two-stream structure with shuffled connections between magnitude and phase estimations. (d) Two-stream structure with independent connections between magnitude and phase estimations.}
		\label{fig:structure}
		\vspace{-8pt}
	\end{figure}
	\renewcommand\arraystretch{0.95}
	\begin{table*}[htbp]
		\centering
		\caption{The experimental results of transforming APNet2 to Vocos on the LibriTTS dataset. ``Direct'' denotes directly estimate the phase spectrum, and ``Atan'' denotes estimating the phase via the Atan function. ``V-L'' and ``A-L'' denotes adopting the loss setups by Vocos and APNet2, respectively.
		}
		\resizebox{0.86\textwidth}{!}{
			\begin{tabular}{c@{\hspace{0.6em}}c@{\hspace{0.6em}}c@{\hspace{0.6em}}c@{\hspace{0.6em}}
					c@{\hspace{0.6em}}c@{\hspace{0.6em}}c@{\hspace{0.6em}}c@{\hspace{0.6em}}
					c@{\hspace{0.6em}}c@{\hspace{0.6em}}c@{\hspace{0.6em}}c@{\hspace{0.6em}}
					c@{\hspace{0.6em}}c@{\hspace{0.6em}}c}
				\toprule
				\rowcolor{gray!20} & &  &  & & &  &   &&   &  & V/UV& Periodicity & Pitch & Param. \\ 
				\rowcolor{gray!20}\multirow{-2}{*}{Ids} &\multirow{-2}{*}{Model} &\multirow{-2}*{Unit} &\multirow{-2}*{Output} &\multirow{-2}*{Loss}  &\multirow{-2}*{Topology}   &\multirow{-2}*{PESQ$\uparrow$}  &\multirow{-2}*{UTMOS$\uparrow$}  &\multirow{-2}*{VISQOL$\uparrow$}&\multirow{-2}*{MCD$\downarrow$}   &\multirow{-2}*{M-STFT$\downarrow$}  & F1$\uparrow$ & RMSE$\downarrow$  & RMSE$\downarrow$  & (M) \\
				
				\midrule
				1& Vocos      & ConvNext v1  & Direct & V-L  & Shared                & \textbf{3.487} &\textbf{3.356}  &\textbf{4.861}&\textbf{3.260} & \textbf{0.921} & \textbf{0.945} & \textbf{0.124}       & \textbf{32.846}     & 13.53  \\ \midrule

				\rowcolor{gray!10}2&      & ConvNext v2 & Atan   & A-L &                 & 1.626 &1.663  &4.397& 6.173 & 1.674 & 0.752 & 0.314 & 254.807 & 31.57 \\
				3&     & ConvNext v1  & Atan   & A-L  &               & 2.194 &2.203  &4.558& 5.215 & 1.371 & 0.893 & 0.194       & 93.412     & 31.53  \\
				\rowcolor{gray!10}4&          & ConvNext v1  & Direct & A-L  &              & 1.529 &1.334  &4.294& 7.183 & 2.162 & 0.697 & 0.365       & 236.895    & 26.54  \\ 
				5&\multirow{-4}{*}{APNet2}  & ConvNext v1  & Direct & V-L  &\multirow{-4}{*}{Separate}                  & 2.556 &1.886  &4.755& 4.232 & 1.123 & 0.905 & 0.193       & 66.027     & 26.54  \\ 
				
				\bottomrule
		\end{tabular}}
		\vspace{-0.35cm}
		\label{tab:apnet2_results}
	\end{table*}
	\vspace{-9pt}
	\renewcommand\arraystretch{0.95}
	\begin{table*}[t]
		\centering
		\caption{The experimental results of different architectural topologies on the LibriTTS dataset. * denotes that the model is only different from Vocos in architectural topology. "R" indicates the number of weight-sharing layers.}
		\resizebox{0.86\textwidth}{!}{
			\begin{tabular}{cccccccccccc}
				\toprule
				\rowcolor{gray!20}  &   &  &   & &&   &  & V/UV& Periodicity & Pitch & Param. \\ 
				
				\rowcolor{gray!20}\multirow{-2}*{Ids} &\multirow{-2}*{Model}  &\multirow{-2}*{Topology}  &\multirow{-2}*{PESQ$\uparrow$}     &\multirow{-2}*{UTMOS$\uparrow$} &\multirow{-2}*{VISQOL$\uparrow$}&\multirow{-2}*{MCD$\downarrow$}   &\multirow{-2}*{M-STFT$\downarrow$}   & F1$\uparrow$ & RMSE$\downarrow$  & RMSE$\downarrow$  & (M) \\
				
				\midrule
				1 & Vocos-D    &   Shared              & \textbf{3.607}  &\textbf{3.461} &\textbf{4.883}& \textbf{3.145} & \textbf{0.897} & \textbf{0.952} & \textbf{0.114}       & 34.694     & 29.73  \\ 
				
				\rowcolor{gray!20}2 & APNet2* &    Separate             & 2.556  &1.886 &4.755& 4.232 & 1.123 & 0.905 & 0.193       & 66.027     & 26.54  \\ 
				3 & APNet2*-SF & Shuffle                & 3.400  &3.344 &4.861& 3.297 & 0.903 & 0.946 & 0.126       & 35.622     & 26.54  \\ 
				\rowcolor{gray!20}4 & APNet2*-PS  & Partially Shared (R=2) & 3.382  &3.182 &4.863& 3.373 & 0.912 & 0.943 & 0.130 & 35.921  & 28.99 \\
				5 & APNet2*-PS  & Partially Shared (R=4) & 3.419  &3.136 &4.862& 3.359 & 0.914 & 0.944 & 0.128       & \textbf{32.628}     & 30.72  \\
				\rowcolor{RoyalBlue!12} 6 & APNet2*-PS  & Partially Shared (R=6) & 3.461  &3.367 &4.873& 3.265 & 0.894 & 0.948 & 0.120 & 31.980 & 31.07  \\
				\bottomrule
			\end{tabular}
            }
		\vspace{-0.35cm}
		\label{tab:modality_results}
	\end{table*}
	\section{Methodology}\label{sec:methodology}
	\label{sec:method}
	\vspace{-6pt}
	\subsection{Rethinking Frequency Vocoders: A Topology Perspective}\label{sec:different-architecture-modality}
	\vspace{-3pt}
	For T-F domain-based neural vocoders, direct or implicit estimation of magnitude and phase is crucial for high-quality audio generation. In this paper, we choose two open-sourced vocoders, namely Vocos{\footnote{https://github.com/gemelo-ai/vocos}} and APNet2{\footnote{https://github.com/redmist328/APNet2}}, which possess relatively similar network modeling units, loss functions, training pipelines, but different in architectural topology, as a typical example to investigate the effective way of joint estimation of MP pairs, and the overall network diagrams are shown in Fig.~{\ref{fig:framework}}(a)-(b), respectively.
	
	We evaluate the generation quality of the two vocoders on two common benchmarks: the LJSpeech~{\cite{ljspeech}} and LibriTTS~{\cite{libritts}} datasets. The former is a single-speaker dataset whose duration is around 24 h, and the latter is a more acoustic-diverse dataset, containing around 585 h duration recorded from 2,456 speakers. As illustrated in Fig.~\ref{fig:pesq_comparison}, Vocos and APNet2 achieve comparable PESQ and UTMOS scores on the LJSpeech dataset. However, APNet2 undergoes severe performance collapse on the more diverse and larger LibriTTS benchmark{\footnote{We provide the training tensorboard of two models at~\url{https://github.com/lingling-dai/Mag_Pha_Tensorboard} for clear illustrations.}}, as shown in id1-id2 of Table~{\ref{tab:apnet2_results}}. To isolate the key factor that led to the phenomenon, we first conduct quantitative experiments by gradually substituting the network unit, output type, and loss functions in APNet2 with those in Vocos. As presented in id3-id5 of Table~\ref{tab:apnet2_results}, after gradually excluding all the possible factors above, we do not observe substantial performance improvement of APNet2, which points to the remaining factor: \textbf{architectural topology}, or more specifically, the information interaction mechanism between magnitude and phase targets. 
	\vspace{-6pt}
	\subsection{Optimizing in the Topological Space}\label{sec:optimizing-from-the-topology-space}
	\vspace{-3pt}
	When inspecting the generation process of MP pairs in Vocos and APNet2, one can observe they belong to Fig.~{\ref{fig:structure}}(a) and (d), respectively. Concretely, in Vocos, magnitude and phase share the main modeling units, and are only independent at the output heads, whereas in APNet2, independent modeling streams are utilized for magnitude and phase, respectively, and no interactions are involved. Recently, in addressing the phase estimation challenges due to its intrinsic wrapping characteristics, feature guidance from the magnitude branch has been proven helpful~\cite{PHASEN,zheng2021interactive}. Therefore, we attempt to optimize the architecture of APNet2 from the topological space. To be specific, in Fig.~{\ref{fig:structure}}(b)-(c), we present two topology variants, which enable more information interaction or weight sharing between the magnitude and phase streams. In Fig.~{\ref{fig:structure}}(b), the magnitude and phase branches share partial feature modeling while leaving separate modeling capabilities for both targets. Moreover, in Fig.~{\ref{fig:structure}}(c), we introduce shuffled operations between the magnitude and phase branches at various layer levels, with information exchange from previous layers while maintaining independent processing. 
	\vspace{-6pt}
	\subsection{Optimizing in the Source Space}\label{sec:source-space}
	\vspace{-3pt}
	Apart from substituting a more effective architectural topology for joint magnitude and phase prediction, we also consider solutions with minimal modifications on model structures for further improvements. Instead of directly posing feature sharing or information interaction mechanism from the topological space, introducing a magnitude-related term that aligns with the target magnitude in the source space seems to alleviate the need for information exchange from the magnitude stream. Additionally, adopting prior knowledge as input features also benefits model convergence and performance improvement \cite{lpcnet,DSPGAN_2022,baenet,liu2024rfwave,hao2021fullsubnet}. 
	
	Motivated by~\cite{freev, rndvoc}, we use the pseudo-inverse mel-spectrogram as an input feature, which serves as an effective representation of range space. Let $\mathbf{S} \in \mathbb{R}^{F_m \times T}$ denote the mel-spectrogram, where $F_m$ is the number of mel bins and $T$ is the number of frames. The pseudo-inverse mel-spectrogram $\mathbf{S}_{\text{prior}} \in \mathbb{R}^{F \times T}$, with $F$ being the number of STFT frequency bins, is computed as:
	\begin{equation}\label{eqn:1}
		\mathbf{S}_{\text{prior}} = \mathbf{W}_{\text{mel}}^{\dag} \mathbf{S},
	\end{equation}
	where $\mathbf{W}_{\text{mel}}^{\dag} \in \mathbb{R}^{F\times F_{m}}$ is the pseudo-inverse of the mel filter bank matrix. This operation reconstructs a coarse linear spectrogram from the mel-spectrogram, providing richer frequency information as input to the vocoder. By taking $\mathbf{S}_{\text{prior}}$ as the input, the model benefits from a more informative initialization for both magnitude and phase prediction, as well as finer guidance for the phase stream.
	\renewcommand\arraystretch{0.90}
	\begin{table*}[]
		\centering
		\caption{The experimental results of making modifications from the source space and output space on the LibriTTS dataset. $\dag$ denotes that the model further adopts the prior source and the MI-RI output type. ``Prior" denotes use pseudo-inverse mel-spectrogram as input.}
		\resizebox{0.90\textwidth}{!}{
			\begin{tabular}{cc
					ccc
					ccc
					ccc
					ccc}
				\toprule
				\rowcolor{gray!10}&     &    &        &    &    & &&   &  & V/UV& Periodicity & Pitch & Param. \\ 
				
				\rowcolor{gray!10}\multirow{-2}*{Ids} &\multirow{-2}*{Model} &\multirow{-2}*{Topology}   &\multirow{-2}*{Source}  & \multirow{-2}*{Output}    &\multirow{-2}*{PESQ$\uparrow$}   &\multirow{-2}*{UTMOS$\uparrow$} &\multirow{-2}*{VISQOL$\uparrow$}&\multirow{-2}*{MCD$\downarrow$}  &\multirow{-2}*{M-STFT$\downarrow $}  & F1$\uparrow $& RMSE$\downarrow $ & RMSE$\downarrow $ & (M) \\
				\midrule
				
				1 &  &                &Raw   & Direct    & 2.556  &1.886 &4.755& 4.232 & 1.123 & 0.905 & 0.193       & 66.027     & 26.54  \\
				\rowcolor{gray!10}2 &                       &                    & Prior  & Direct    & 3.197  &2.987 &4.846& 3.500 & 0.944 & 0.936 & 0.138       & 42.139     & 29.50\\
				\cmidrule{4-14}
				3 &  &                                    &     & MI-RI     &  3.436  &3.202 &4.869& 3.367 & 0.929 & 0.948 & 0.123      & 30.394     &\\
				\rowcolor{gray!10}4 &\multirow{-4}{*}{APNet2*}                 &\multirow{-4}{*}{Separate}                                      &\multirow{-2}{*}{Raw}        &  MB       &3.436  &3.202 &4.869& 3.368 & 0.929 & 0.948 & 0.123 & 30.348     & \multirow{-2}{*}{26.80} \\ \midrule
				

				5 & Vocos$\dag$-D                    & Shared                 & Prior & MI-RI  & 3.576  &3.369 &4.880& 3.247 & 0.897 & 0.948 & 0.121       & 35.721     & 32.35  \\
				\rowcolor{RoyalBlue!5}6 & APNet2$\dag$                     & Separate               & Prior & MI-RI  & 3.569  &\textbf{3.410} &\textbf{4.896}& \textbf{3.199} & \textbf{0.883} & 0.948 & 0.117       & 29.580     & 29.76  \\ 
				7 &APNet2$\dag$-SF                  & Shuffle                & Prior & MI-RI  & 3.510  &3.325 &4.873& 3.282 & 0.900 & 0.949 & 0.120       & 34.863     & 29.76  \\
				\rowcolor{RoyalBlue!15}8 & APNet2$\dag$-PS                  & Partially Shared (R=2) & Prior & MI-RI  & \textbf{3.610}  &\textbf{3.410} &4.886& 3.219 & \textbf{0.883} & 0.950 & \textbf{0.114}       & 30.602     & 30.95  \\
				9 & APNet2$\dag$-PS                  & Partially Shared (R=4) & Prior & MI-RI  & 3.573  &3.396 &4.887& 3.239 & 0.884 & \textbf{0.951} & 0.115       & \textbf{27.686}     & 32.90  \\
				\rowcolor{RoyalBlue!25}10 & APNet2$\dag$-PS                  & Partially Shared (R=6) & Prior & MI-RI  & 3.511  &3.314 &4.872& 3.289 & 0.900 & 0.948 & 0.120       & 31.799     & 33.46  \\
				\bottomrule
		\end{tabular}
        }
		\vspace{-6pt}
		\label{tab:proposed_results}
	\end{table*}
	\vspace{-6pt}
	\subsection{Optimizing in the Output Space}\label{sec:output-space}
	\vspace{-3pt}
	In T-F domain-based neural vocoders, the magnitude is usually relatively easier for prediction due to its clear structural characteristic, while phase representation remains an active and open area of research. For instance, a mainstream of models predicts phase with the Atan operation on network outputs~\cite{CRM}. More recently, Vocos~\cite{vocos} proposed to estimate the phase directly and apply the cosine and sine calculations for RI estimation. However, they still face the following challenges: \\
	\noindent 1) \textbf{Wrapping issue}: The phase spectrogram is inherently wrapped within the range of $(-\pi, \pi]$, leading to discontinuities that can complicate the learning process for neural networks. \\
	\noindent 2) \textbf{Multiple-solution problem}: $\text{Atan}(k\tilde{I}, k\tilde{R})$ or $\theta+2h\pi$ with different values of $k \in \mathbb{R}$ or $h \in \mathbb{Z}$ target to one final result but with different network outputs, making the optimization more difficult. \\
	\noindent 3) \textbf{Uncertain coherence with magnitude}: The generation of phase is less dependent on the generated magnitude for architectural topologies with non-shared parts, which may lead to inconsistent results when reconstructing the waveform.
	
	As illustrated in Fig.~\ref{fig:framework}(d), we propose a simple yet effective modification on APNet2 by enforcing the phase components to 
	estimate partial magnitude components, which facilitates the joint parameter update in the backpropagation process. Specifically, we utilize the estimated real component $\tilde{R}$ and imaginary component $\tilde{I}$ to add a magnitude-related item $\hat{A}_p$ for the final magnitude prediction:
	\begin{equation}\label{eqn:2}
		\hat{A}=\alpha \hat{A}_M+\left( 1-\alpha \right) \hat{A}_p, 
	\end{equation}
	\begin{equation}\label{eqn:3}
		\hat{A}_{p}=\sqrt{\tilde{R}^{2} + \tilde{I}^{2}}, 
	\end{equation}
	where $\alpha$ is a trainable weighting hyper-parameter. During backpropagation, taking the MSE magnitude-loss function as an example, the gradient flow of the network parameter $\Phi$ is computed as:
	\begin{equation}
		\label{eqn:4}
		\frac{\partial \left( \hat{A}-A \right) ^2}{\partial \Phi}=2\left( \hat{A}-A \right) \left[ \frac{\partial \left( \alpha \hat{A}_M \right)}{\partial \Phi _M}+\frac{\partial \left( \left( 1-\alpha \right) \hat{A}_{P} \right)}{\partial \Phi _{P}} \right], 
	\end{equation}
	where the magnitude-related rather than phase-related quantities are introduced to drive the overall parameter update, thereby alleviating the optimization challenges brought by the incorrect phase estimation. Furthermore, merely the magnitude loss term allows the whole network parameters to be updated, which indirectly enhances coherence between phase and magnitude generation. Besides, the value of RI components is also constrained by magnitude under such optimization, thereby relieving the problem of multiple solutions.
	\vspace{-8pt}
	\section{Experimental Setup}
	\vspace{-3pt}
	The experiments are based on LibriTTS benchmark. Following the data split principle in BigVGAN~{\cite{bigvgan}}, we utilize all the training subsets $\left\{\textit{train-clean-100}, \textit{train-clean-360}, \textit{train-other-500}\right\}$ for training, and $\left\{\textit{test-clean}, \textit{test-other}\right\}$ subsets for evaluation. We employ official implementations of Vocos and APNet2. Note that the performance may be different from the official checkpoints as we remove the amplitude augmentation during training and align the learning rate to 5e-4 for fair comparisons. For objective evaluations, we employ eight objective metrics: perceptual evaluation of speech quality (PESQ)~\cite{WB-PESQ}, UTMOS~\cite{UTMOS}, Virtual Speech Quality Objective Listener (VISQOL)~\cite{hines2015visqol}, mel-cepstral distortion (MCD)~\cite{mcd}, multi-resolution STFT loss (M-STFT)~\cite{mstft}, Periodicity Root Mean Square Error (RMSE), V/UV F1 score, and pitch RMSE~\cite{f1_score}. 
	\vspace{-6pt}
	\section{Results and Analysis}
	\vspace{-6pt}
	\subsection{Evaluation on Different Architectural Topologies}
	To verify the effectiveness of the feature-sharing mechanism introduced between the magnitude and phase streams in the topological space, we conduct comprehensive experiments on models of different architectural topologies with close parameter sizes. We denote the partially shared structure as APNet2*-PS and the two-stream structure with shuffled connections as APNet2*-SF, and a large version of Vocos as Vocos-D. As presented in Table~\ref{tab:modality_results}, one can observe that APNet2*-PS and APNet2*-SF significantly improve the performance of APNet2, revealing that the choice of architectural topologies can significantly impact the performance of frequency-domain neural vocoders, as it influences the way magnitude and phase information are processed and integrated during audio generation. 
	\vspace{-12pt}
	\subsection{Evaluation of Source Space and Output Space}
	\vspace{-3pt}
	In Table~\ref{tab:proposed_results}, we first provide ablation studies on the effectiveness of modifications in the source space and the output space, respectively. By introducing the pseudo-inverse mel-spectrogram in the source space, the performance of APNet2* improves significantly, with PESQ increasing from 2.556 to 3.197, indicating the effectiveness of applying magnitude initialization. Additionally, after replacing the MI-RI output format, the performance of APNet2* also improves. As we separate the magnitude item from the magnitude branch (MB), one can observe only a very slight difference from the final results, revealing that the magnitude component $\hat{A}_P$ from phase branch attributes merely no influence in the forward process of audio generation and our proposed MI-RI contributes positively to the parameter update in the backpropagation process when the magnitude and phase streams involves insufficient information interaction. We further apply our proposed method in source space and output space to all the above-mentioned architectural topologies. As presented in Table~\ref{tab:proposed_results}, our proposed method effectively improves the performance of models with less effective information interaction between magnitude and phase streams, and also minimizes their performance disparities with Vocos-D. 
	
	\begin{figure}[htb]
		\centering
		\includegraphics[width=0.44\textwidth]{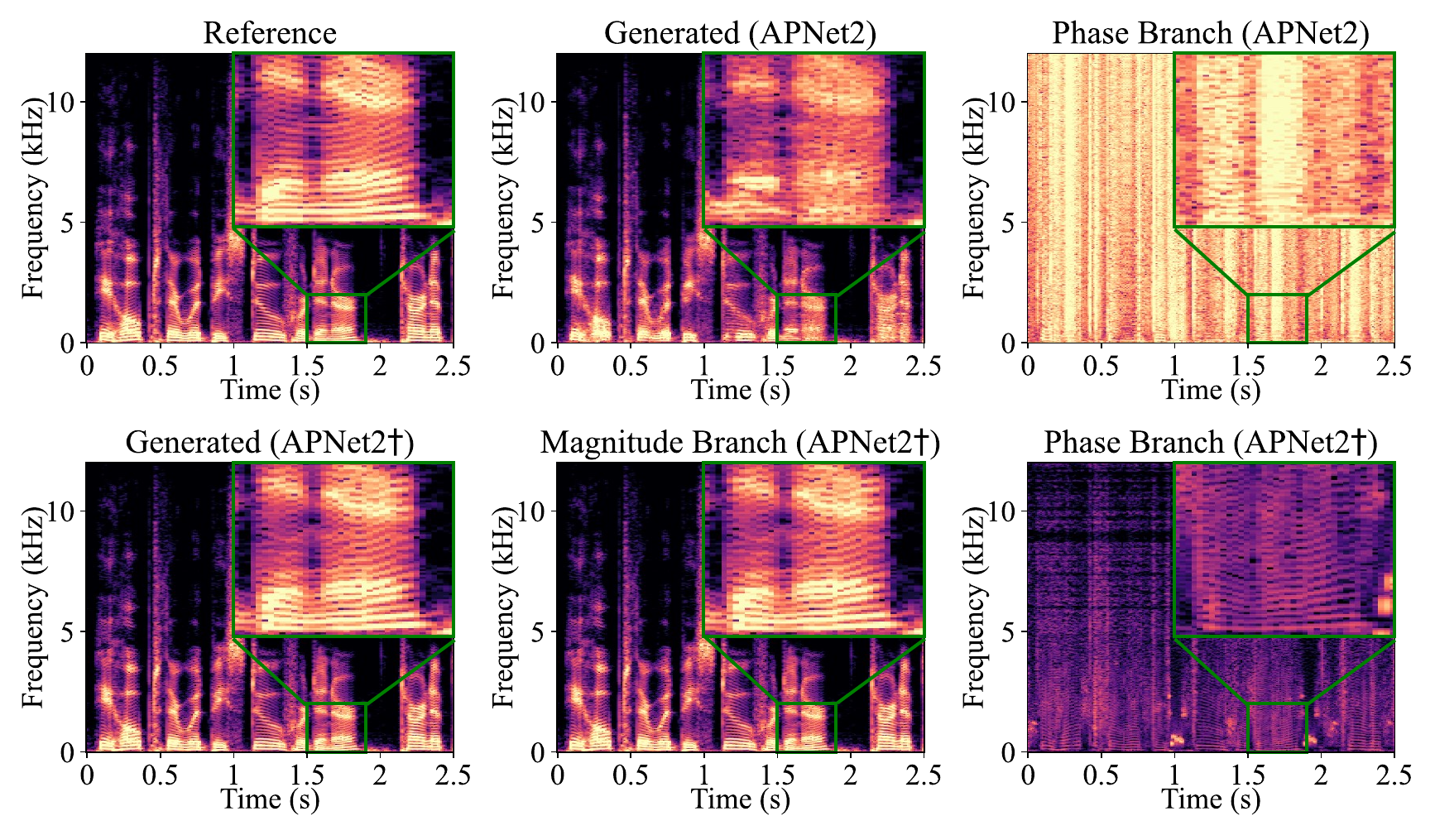}
		\caption{Spectrogram visualization of the reference, generated audio signal by APNet2$\dag$ and APNet2, and the intermediate results of APNet2$\dag$ and APNet2, respectively.}
		\vspace{-8pt}
		\label{fig:visualization}
	\end{figure}
	\vspace{-6pt}
	\vspace{-6pt}
	\subsection{Visualization of the Intermediate Results}
	\vspace{-3pt}
	In Fig.~\ref{fig:visualization}, we present the spectrogram visualization of the reference, the generated audio signal from APNet2$\dag$ and APNet2, and the intermediate results of APNet2$\dag$ and APNet2, where the magnitude components are derived separately from the magnitude branch and the phase branch. It can be observed that the magnitude component $\hat{A}_P$ from the phase branch in APNet2$\dag$ exhibits clear harmonic structures that align with the spectrogram of the generated audio, whereas in APNet2, the magnitude component derived from the phase branch does not exhibit clear structural details. This finding further demonstrates the effective improvement of our proposed method in enhancing the coherence between magnitude and phase. Additionally, the magnitude component $\hat{A}_M$ from the magnitude branch illustrates no significant difference compared with the final generated spectrogram, further verifying that our proposed MI-RI does not affect the forward process. 
	\vspace{-10pt}
	\section{Conclusions}
	\label{sec:conclusion}
	\vspace{-4pt}
	In this paper, we revisit the joint magnitude and phase estimation mechanism on two typical neural vocoders: Vocos and APNet2. To bridge their performance gaps, we propose simple yet effective modifications in three distinct spaces, including the topological space, the source space, and the output space. Specifically, we propose to enhance the information interaction in the topological space, introduce prior knowledge to the source space, and optimize the backpropagation process with an improved output format in the output space. Experimental results demonstrate that our proposed method effectively aligns the performance of APNet2 with Vocos, both jointly and exclusively.

	\footnotesize
	\bibliographystyle{IEEEbib}
	\bibliography{refs}
	
\end{document}